\documentstyle[aps,preprint]{revtex}

\begin{document}

\draft

\title{Kondo effect in AC transport through Quantum Dots }
\author{Rosa L\'opez$^{1}$, Ram\'on Aguado$^{1,2}$, Gloria Platero$^{1}$ and
Carlos Tejedor$^{3}$}
\address{$^1$ Teor\'{\i}a de la Materia Condensada,
Instituto de Ciencia de Materiales (CSIC) Cantoblanco,28049
Madrid, Spain.}
\address{$^2$Department of Applied Physics,
Delft University of Technology,
Lorentzweg 1, 2628CJ, Delft, The Netherlands.}
\address{$^3$ F\'{\i}sica Te\'orica
de la Materia Condensada,
Universidad Aut\'onoma de Madrid, Cantoblanco,28049
Madrid, Spain.}

\date{\today}
\maketitle

\begin{abstract}

A theory of the Kondo effect in quantum dots at zero temperature
in the presence of arbitrarily
intense AC potentials is presented. We generalize the Friedel-Langreth sum
rule to take care of charge conservation and propose a consistent procedure
to study a time dependent Anderson Hamiltonian.
The effect of the AC potential on both the quantum dot density of states and
the linear conductance shows the importance of using a theory which
describes intradot finite interaction and nonperturbative effects.

\end{abstract}



The competition between electron-electron interaction and quantum-mechanical
hybridization between the delocalized electrons in a non-magnetic metal and
the unpaired electrons of a magnetic impurity leads to
the Kondo effect \cite{Hew}.
It has been predicted \cite{Ng,Gla,Kaw,Mei,Hers,Lev,Win,Wan} 
that the transport at low temperatures through a
single level coupled to two reservoirs containing Fermi 
liquids \cite{chann}, is governed by a Kondo-like
singularity existing in the quasiparticle density of states
(DOS) at the Fermi level $\epsilon_F$ of the reservoirs.
This has been experimentally performed recently\cite{Gold,Leo} by means of a
quantum dot (QD) coupled to two leads by tunneling barriers. When {\it i)} 
the QD is small to have well separated levels, {\it ii)} the tunneling rate 
is such that the broadening of QD levels is typically one order of magnitude 
smaller than the QD charging energy and {\it iii)} the number
of electrons within the QD is odd, then, there is an unpaired electron which 
is free to form a singlet state with the electrons in the leads. This is the
physics behind of the Kondo effect that is well described by
the low energy excitations of the Anderson Hamiltonian for the unpaired
electron while all the other electrons are effectively described by a 
potential of the QD coupled to the leads.
The excitations around $\epsilon_F$ are quasiparticles having a lifetime 
determined only by single particle properties without any many-body 
contribution. One electron at $\epsilon_F$, becomes scattered
by the dot suffering a phase shift which is proportional to the exact
displaced charge $\delta n$.
As a consequence of this Friedel-Langreth (FL) theorem \cite{Lan} the
conductance takes the value ${\cal G}=sin^2(\pi \delta 
n)2e^{2}/h$\cite{Ng,Gla}.
The interest of the Kondo physics in QD resides in
the possibility of externally controlling parameters as the on-site
interaction $U$, the QD level $\epsilon_0$ or the broadening $\Gamma$ of
the DOS, due to the dot-leads coupling \cite{Gold,Leo}.
It is also possible to study the Kondo singularity itself by
applying a bias in order to mimic the QD DOS by the $dI/dV$
curve \cite{Lev,Gold,Leo}.

In this letter we address the following question: since the 
quasiparticles with large lifetime are those at $\epsilon _F$, what
happens to the Kondo problem at zero temperature when there is a
mixing of states with
energies differing from $\epsilon _F$ by a finite amount?
In the case of QD, this can be actually made by means of an AC gate
voltage\cite{Hett,Ng2,Sch,Win2}. The problem is very interesting and
far from trivial because the frequency range experimentally accessible
(from $\sim 100$ MHz to $20$ GHz)\cite{Leo2} is roughly that of the Kondo
temperature $T_K=\sqrt {U\Gamma } exp \{-\pi[|\epsilon _F -\epsilon _0|
(U+\epsilon _0)]/\Gamma U\}$ (from few mK to $\sim 1$K)\cite{Leo}.

The aim of our work is to describe the QD DOS within a framework where
charge conservation is taken into account in order to describe properly
the Kondo singularity.
Therefore, our first task is to generalize the FL sum rule to the case of
a time dependent situation.
The Anderson Hamiltonian is solvable through Bethe ansatz \cite{Beth}
or Quantum Monte Carlo methods \cite{MC} but a reliable and simple method
to obtain dynamical properties at low temperatures in the whole range of
interactions ($U/\Gamma$) is not available.
Previous efforts concentrated in the $U\rightarrow\infty$ limit where a
Non-Crossing Approximation \cite{Mei,Hett,Win2} can be made for high and
intermediate temperatures. However, such approximations do not give
the exact local Fermi-liquid properties when $T\rightarrow 0$ and
cannot describe the transition from the
weak-correlation to the strong-correlation regime.
On the other hand, finite $U$ perturbation theory \cite{Hers,Yos,Hor}
describes properly the symmetric case but presents clear anomalies
away from this special situation. Our method is an extension of those
\cite{Lev,Kot} developed in the static case where an effective selfenergy
eliminates consistently the problems of the perturbative approaches.

The time dependent Anderson Hamiltonian reads:
\begin{eqnarray}
H & = & \sum _{ k\in \{ L,R \}, \sigma} \epsilon _k c^\dagger _{ k, \sigma}
c_{ k, \sigma} + \sum _\sigma \left( \epsilon _\sigma +
V_{AC}cos \omega_0 t \right) d^\dagger_\sigma d_\sigma \nonumber \\
& + & \sum _{k\in \{L,R\},\sigma }V_k (c^\dagger _{k, \sigma} d_\sigma
+ d^\dagger_\sigma c_{k, \sigma}) + U d^\dagger_\uparrow d_\uparrow
d^\dagger_\downarrow d_\downarrow.
\end{eqnarray}
$d^\dagger_\sigma$ creates an electron with spin $\sigma$ in the dot
while $c^\dagger _{k, \sigma}$ creates it in the lead with energy $\epsilon
_k $ ($k$ labels the rest of quantum numbers). The AC voltage with intensity
$V_{AC}$ and frequency $\omega_0 $
modulates in time the relative position of the QD level $\epsilon _\sigma $
with respect to $\epsilon_F$. An eventual breakdown of the spin
degeneracy would be represented by $\epsilon _\sigma\neq \epsilon 
_{-{\sigma}}$.
The coupling $V_{k}$ between the QD and the leads
produces a broadening
$\Gamma^{L(R)}=-2 Im [ \Sigma ^{L(R)} _{sp} (\epsilon +i \delta )]=
2\pi\sum_{k\in L(R)}|V_{k} |^2\delta(\epsilon-\epsilon_{k})$.
The dc linear conductance at zero temperature deduced from a nonequilibrium
Keldysh technique is\cite{Jau}.
\begin{eqnarray}
{\cal G} &=&\frac{e^2}{\hbar}
\frac{\Gamma^{L}(\epsilon_F)\Gamma^{R}(\epsilon_F)}{\Gamma(\epsilon_F)}
\sum_\sigma \rho_\sigma(\epsilon_F)
\label{conductance}
\end{eqnarray}
where $\Gamma=\Gamma^{L}+\Gamma^{R}$.
${\cal G}$ is related to the time averaged DOS at the QD
$\rho_\sigma(\epsilon)=-Im \langle A_{\sigma}
(\epsilon)\rangle /\pi$ obtained from 
\begin{eqnarray}
\langle
A_{\sigma}(\epsilon)\rangle&=&\frac{\omega_0}{2\pi}\int_{0}^{\frac{2\pi}{
\omega_0}}dt\int_{-\infty}^{t}dt_{1}G^{r}_{\sigma}(t,t_{1})
e^{i\epsilon(t-t_{1})}
\label{spectralf}
\end{eqnarray}
$G^{r}_{\sigma}(t,t_{1})$ being the QD retarded Green's function.
Coherent tunneling of one electron at $\epsilon _F$ through the QD with
AC produces a phase shift of its wave function which is related, according
to the FL rule, to the displaced charge for each spin which is related with
the QD occupation ($\langle n_\sigma\rangle$) and with
the number of lead electrons with ($\langle n^{l}_\sigma\rangle$) and
without ($\langle n^{l}_{0,\sigma}\rangle$) coupling to the QD:
\begin{eqnarray}
\eta _\sigma (\epsilon_{F})=\pi (\langle n_\sigma\rangle + \langle
n^{l}_\sigma\rangle-\langle n^{l}_{0,\sigma}\rangle).
\label{FL}
\end{eqnarray}

The phase shift is related with $\langle A_{\sigma}(\epsilon)\rangle $ 
through the eigenvalues of the scattering ${\cal T}_\sigma$ matrix
\begin{eqnarray}
e^{2i\eta _\sigma (\epsilon )}=1+{\cal T}_\sigma (\epsilon )
=1-i\sqrt{\Gamma _{L}\Gamma _{R}}
\langle A_{\sigma}(\epsilon)\rangle.
\label{phaseshift}
\end{eqnarray}
The displaced charge in Eq.(\ref{FL}) can be obtained from the
retarded Green's function as follows:
{\em i)} $\langle  n^{l}_\sigma \rangle = -
Im\langle G_{k,\sigma}^{<}(t,t)\rangle$ and $\langle n_\sigma \rangle =
-Im\langle G_{\sigma}^{<}(t,t)\rangle $.
{\em ii)} $G^<$ is related to the retarded Green's function built up
using $G^{r}_{\sigma}(t,t_{1})= G^{r}_{\sigma}(t-t_{1}) \!
exp \left[ -\frac{i}{\hbar } \int_{t}^{t_{1}}d\tau V_{AC}cos \omega_0\tau
\right] $, and {\em iii)} application of the operational rules
given by Langreth \cite{Lan2} for Keldysh contour integration. This process,
together with Eqs.(\ref{FL}) and (\ref{phaseshift}), gives
\begin{eqnarray}
& & Im \; ln \left( \sum_{m=-\infty}^{\infty}J^2_m(\beta
)G^r_\sigma (\epsilon_F- m\hbar\omega _0) \right) =
\nonumber \\
& & -Im \sum_{m=-\infty}^{\infty} J^2_m(\beta) \int_{-\infty}^{\epsilon_F}
d\epsilon \;
\frac{\Gamma ^{L}(\epsilon)+\Gamma ^{R}(\epsilon)}
{\Gamma (\epsilon- m\hbar\omega _0)}
G^r_\sigma (\epsilon- m\hbar\omega _0)
\nonumber \\
& & \times \left( 1-\sum _{s=-\infty }^\infty J_s^2(\beta) \;
\sum _k \frac{|V_{k}|^2}{[\epsilon-\epsilon _k-(m-s)\hbar\omega _0]^2}
\right)
\label{newFL}
\end{eqnarray}
where $G^r_\sigma (\epsilon )$ is the Fourier transform of the static
retarded Green's function and $J_m$ the Bessel function of order $m$ 
and argument  $\beta = V_{AC}/\hbar\omega _0$.
Eq.(\ref{newFL}) reduces to the usual FL rule \cite{Lan} for $V_{AC} 
\rightarrow 0$. It establishes the charge neutrality constriction that must be
verified by any approximation made to compute the
retarded Green's function. This result implies a series of important
consequences on the Kondo physics of the QD.

Some approximations are made to compute the retarded Green's function
\begin{eqnarray}
G_\sigma^{r}(\epsilon)=\left[ \epsilon-\epsilon_\sigma 
- \Sigma_{sp}(\epsilon) -\Sigma_\sigma(\epsilon) \right] ^{-1}.
\end{eqnarray}
Correlation effects are included in the selfenergy $\Sigma_\sigma(\epsilon)$.
In order to obtain $\Sigma_\sigma(\epsilon)$, we use an approximation which,
in the static case\cite{Lev,Kot},
eliminates the pathologies of standard perturbation procedures.
$\Sigma_\sigma(\epsilon)$ is calculated by an interpolation that gives
correctly the limits $U/\Gamma\rightarrow 0$ and $\Gamma/U\rightarrow 0$
and has good analytic properties both in $\epsilon\rightarrow \epsilon _F$ 
and $\epsilon \rightarrow \pm \infty$ \cite{Lev,Kot}:
\begin{eqnarray}
& \Sigma_\sigma(\epsilon)=U\langle n_{-\sigma} \rangle
+\frac{\langle n_{-\sigma}\rangle(1-\langle n_{-\sigma}\rangle)}
{\langle n_{0,-\sigma}\rangle (1-\langle n_{0,-\sigma}\rangle)}
\Sigma_\sigma^{(2)}(\epsilon) \nonumber \\
& \times \left(1-
\frac{[(1-\langle n_{-\sigma}\rangle)U+\epsilon_\sigma+\mu_\sigma]}
{\langle n_{0,-\sigma}\rangle
(1-\langle n_{0,-\sigma} \rangle)U^2}
\Sigma_\sigma^{(2)}(\epsilon) \right)^{-1}.
\label{self}
\end{eqnarray}
$\langle n_{0,\sigma} \rangle $ is the mean field (MF) QD occupation 
and $\Sigma_\sigma^{(2)}(\epsilon)$ is the second order 
selfenergy used in previous finite U calculations \cite{Hers,Lev,Kot,Yos,Hor}.
$\mu _ \sigma $ is a shift of the QD level to an effective value 
determined selfconsistently
together with $\langle n_\sigma\rangle$ with the constriction of
charge neutrality given by Eq.(\ref{newFL}).
Once these parameters are determined, 
spectral and transport properties are calculated
from $G^r_\sigma (\epsilon )$.

Hereafter we consider
$\Gamma ^L=\Gamma^R=\Gamma$, $\epsilon_\sigma=\epsilon_{-\sigma}
=\epsilon _0$ and $\epsilon_F=0$.
All our calculations are performed with parameters in the range of the
experiments already available\cite{Gold,Leo} in order to look for
predictions on possible future experiments.
First of all we analyze the effect of the AC field on
the time averaged DOS that can be measured by means of
the $dI/dV$ curve of the QD as reported in the static case \cite{Gold,Leo}.
Fig. \ref{fig:1} shows the DOS for a case with $\epsilon _0 \neq -U/2$.
Results obtained from both Eq.(\ref{self}) for
$\Sigma_\sigma(\epsilon)$ (continuous lines) and
for the selfenergy up to second order (dashed lines) are shown.
The selfconsistency for the selfenergy enhances the particle-hole asymmetry in
both the static (Fig. \ref{fig:1}a) and high frequency (Fig. \ref{fig:1}d) 
cases.
The lowest frequency (Fig. \ref{fig:1}b) corresponds to an adiabatic
modulation of the level ($\hbar\omega_0<\Gamma$) which quenches
the Kondo effect on the time averaged DOS. The two small peaks of
the DOS correspond to the MF (lateral peaks) solutions separated
by the extreme values of the harmonic potential.
Fig. \ref{fig:1}c corresponds to the case $\beta=2$ where the symmetric shape
is recovered due to a transfer of spectral weight
from the Kondo peak to the MF peaks and their satellites.
Fig. \ref{fig:1}d shows the richest configuration with a strong asymmetry.
This is better observed in Fig. \ref{fig:2}
where the evolution of the DOS with $V_{AC}$ is analyzed.
In the case of $V_{AC}=0.4$, one observes three types of peaks: the
ones labeled as $a$ correspond to the Kondo peak and its AC replicas. Those
labeled as $b$ correspond to the low-energy MF peak for $V_{AC}=0$
and its AC replicas. These peaks evolve to higher energies  and become
narrower with increasing $\beta $. Finally, the peaks labeled as $c$
correspond to the high-energy MF peak for $V_{AC}=0$ and its AC 
replicas.
Once again, the evolution with $\beta $ is similar to that of peaks
$b$. Moreover, with increasing intensity, we observe a decrease of both the
height and the width of the peak at the Fermi level. These effects are a
consequence of the appearance of increasing new peaks and the necessary
conservation of the total spectral weight. The reduction of the height
affects the conductance as discussed below while that of the width
implies a change in $T_K$. The behavior of the spectrum is not
trivial ({\it e.g.} occupations change with $\beta$),
and clearly different to that obtained in a single particle
picture of AC-assisted tunneling, where only simple replicas
of the static DOS would appear.

The enormous changes in the DOS when the occupation varies affect strongly
to the conductance. Therefore, the effect of the AC potential on
${\cal G}$ can not be described now by a single-particle model of AC
assisted tunneling through a discrete state\cite{Tien}. Fig \ref{fig:3} 
shows the AC effects on ${\cal G}$. 
At zero temperature and in the absence of AC potential\cite{Ng,Gla,Kaw}
${\cal G}$ has a broad peak of width given by $U$ centered around $\epsilon _0
=-U/2$ and reaches the maximum value of $2e^2/h$\cite{note1}. Fig \ref{fig:3} 
shows a reduction of the central peak (still centered
around $\epsilon _0=-U/2$) and the appearance of symmetrically
located satellites due to the AC. The position of the conductance satellites is
inferred from the following argument: the DOS (left inset of
Fig \ref{fig:3}) has a peak when the renormalized level 
of the QD is equal to $\hbar\omega_0$. 
It has an AC satellite at $\epsilon _F$ (dashed line). This gives a large
contribution to the conductance. For low $\beta $, the level renormalization
is small so that the peak in ${\cal G}$ is located at $\epsilon _0-\hbar
\omega _0 \simeq 0$. Since the conductance is symmetric with respect to $-U/2$,
that satellite has a companion at $\epsilon _0 \simeq -U-\hbar \omega _0$.
So, the separation between satellites is $U+2\hbar \omega _0$ as
observed in Fig. \ref{fig:3} for $V_{AC}=0.2$. When the AC
intensity increases, the QD level renormalization 
increases and the reasoning, still exactly valid for the renormalized level, 
only gives approximately the separation in $\epsilon _0$ between satellites.
In the same range of values for $U$ and $\Gamma $, the results are similar
 for $\hbar \omega _0=U/2$, $U/4$ and $3U/20$.
It must be stressed that whereas the satellites in ${\cal G}$ mark
the AC-assisted transition from the empty orbital regime to the mixed
valence regime, (inset at the left side of Fig \ref{fig:3})
the main broad peak corresponds to tunneling in the Kondo regime
(inset at the right side of Fig \ref{fig:3}).
This behavior in the conductance is different from  the one
predicted by a MF theory, where satellites at $-\hbar \omega _0$
and $\hbar \omega _0-U$  also appear. The lack of these peaks
is due to the different nature of the DOS when passing
from the Kondo regime to the empty orbital regime (single peaked DOS
centered at $\epsilon_0$ ) through the mixed valence
one where the DOS has a single peak centered in a strongly renormalized 
position\cite{Hew,Beth}.
Our general result which depends on U
contrasts with the results of Ref. \cite{Hett}.
They used a model with $U \rightarrow \infty$ and found peaks of half-width
$\Gamma$ in the conductance separated by $\hbar\omega_0$, similar
to previously reported MF AC-assisted tunneling.
Another very interesting result drawn from Fig \ref{fig:3} is the non linear 
reduction of the central peak of the conductance when $V_{AC}$ increases. 
Since the whole behavior of ${\cal G}$ with $\omega _0$ is rather 
different to the linear dependence expected from a MF model,
our result constitutes a proof of the importance of a finite $U$ charge
conserving theory for the AC Kondo effect.

Finally, we pay attention to the effect of the AC potential
on the occupation of the QD level. We show in Fig \ref{fig:4}
the charge in the dot in terms of $\beta $. Results are given for two
different values of $\epsilon _0$ corresponding to initial occupations clearly
above and below 1/2 (per spin) respectively. For increasing $\beta $, 
$\langle n_\sigma \rangle$ tends to 1/2 in a damped oscillatory way. 
For $\beta $ such that $J_0(\beta )\rightarrow 0$, the peak at 
$\epsilon _0$ in the DOS is quenched and its spectral weight 
transferred to the AC satellites. With the chosen parameters half
of the satellites are above/below zero (with the
same intensities), so that the QD occupation tends to 1/2.

In conclusion, we have studied the AC Kondo effect in QD. We extend 
the FL sum rule to consider finite U as well as charge conservation.
Our main findings are: {\it i)} changes in the shape and position of
the peaks in the DOS in the presence of the AC potential.
{\it ii)} variation of the DOS at $\epsilon_F$ and, consequently,
of the linear conductance which has new features with respect to
the usual MF theory. These are finite-U many-body effects.
With regard to the possibility of experimentally observing the features 
described here, narrowing of peaks at the DOS due to AC effects are going to 
reduce $T_K$. However, there is still margin for the measurements: 
for instance, in the mixed valence regime,  
$T_K \sim \Gamma \sim 1K$ in available experiments\cite{Leo} (without AC) 
while temperatures significantly lower can be routinely achieved. 

We acknowledge A.P. Jauho, L. Glazman, L.P. Kouwenhoven, A. Martin-Rodero
and J.J. Palacios for helpful discussions.
Work supported in part by the MEC of Spain under contracts
PB96-0875 and PB96-0085 and by the EU via contract FMRX-CT98-0180.

\begin{figure}
\caption{Time averaged DOS with $V_{AC}=0.2$ for the asymmetric configuration
$U=0.2$, $\epsilon _0=-0.15$ and $\Gamma=0.025$. a) without AC, b) $\hbar
\omega _0=0.01$, c) $\hbar \omega _0=0.1$, 
d) $\hbar\omega_0=0.35$. Solid lines are calculated with the
interpolation procedure for the selfenergy $\Sigma_\sigma(\epsilon)$ (Eq. (
\ref{self})) and dotted lines with the selfenergy up to second order.}
\label{fig:1}
\end{figure}

\begin{figure}
\caption{Time averaged DOS with $\hbar \omega _0=0.35$ for the asymmetric 
configuration $U=0.2$, $\epsilon_0=-0.15$ and $\Gamma=0.025$ (as in Fig. 
\ref{fig:1}) and different intensities $V_{AC}$. The labels {\it a,b,c} 
are discussed in the text.}
\label{fig:2}
\end{figure}

\begin{figure}
\caption{Linear Conductance with
$\hbar \omega _0=0.35$, $U=0.2$ and $\Gamma=0.025$ for $V_{AC}=0$ (continuous
line), $V_{AC}=0.2$ (dotted line) and $V_{AC}=0.4$ (dashed line). The insets
show the DOS with (dashed lines) and without (continuous lines) AC at the
empty orbital (left side, $\epsilon _0=0.35$) and Kondo (right side, 
$\epsilon _0=-0.1$) regimes.}
\label{fig:3}
\end{figure}

\begin{figure}
\caption{Occupation $\langle n_\sigma \rangle$ of the QD as a function of
$\beta $ for $U=0.2$ and $\hbar \omega _0=0.35$. Results with the
selfenergy up to second order (dashed lines) and with Eq.(\ref{self}) 
(continuous lines) are shown for $\epsilon _0=-0.15$ (curves above)
and $\epsilon _0=-0.05$ (curves below).
}
\label{fig:4}
\end{figure}



\begin{thebibliography}{12}
\bibitem{note1} When $T$ 
increases, the center of the peak is quenched and two peaks, 
at $\epsilon_0=-U$ and $\epsilon_0=0$, become apparent. They 
would correspond to the Coulomb Blockade peaks experimentally 
observed in QD for $T>T_K$\cite{Gold,Leo}.
\end{thebibliography}
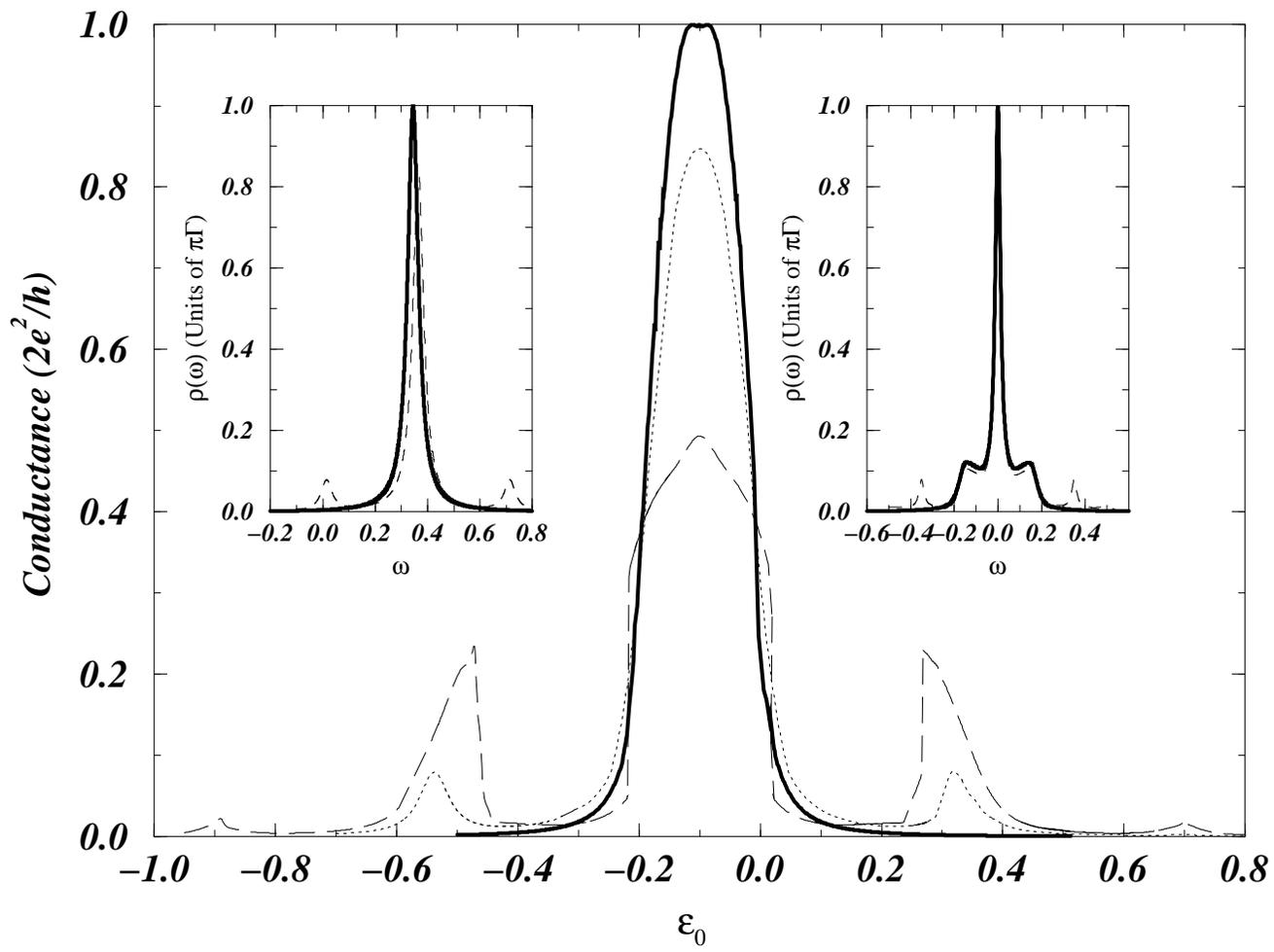
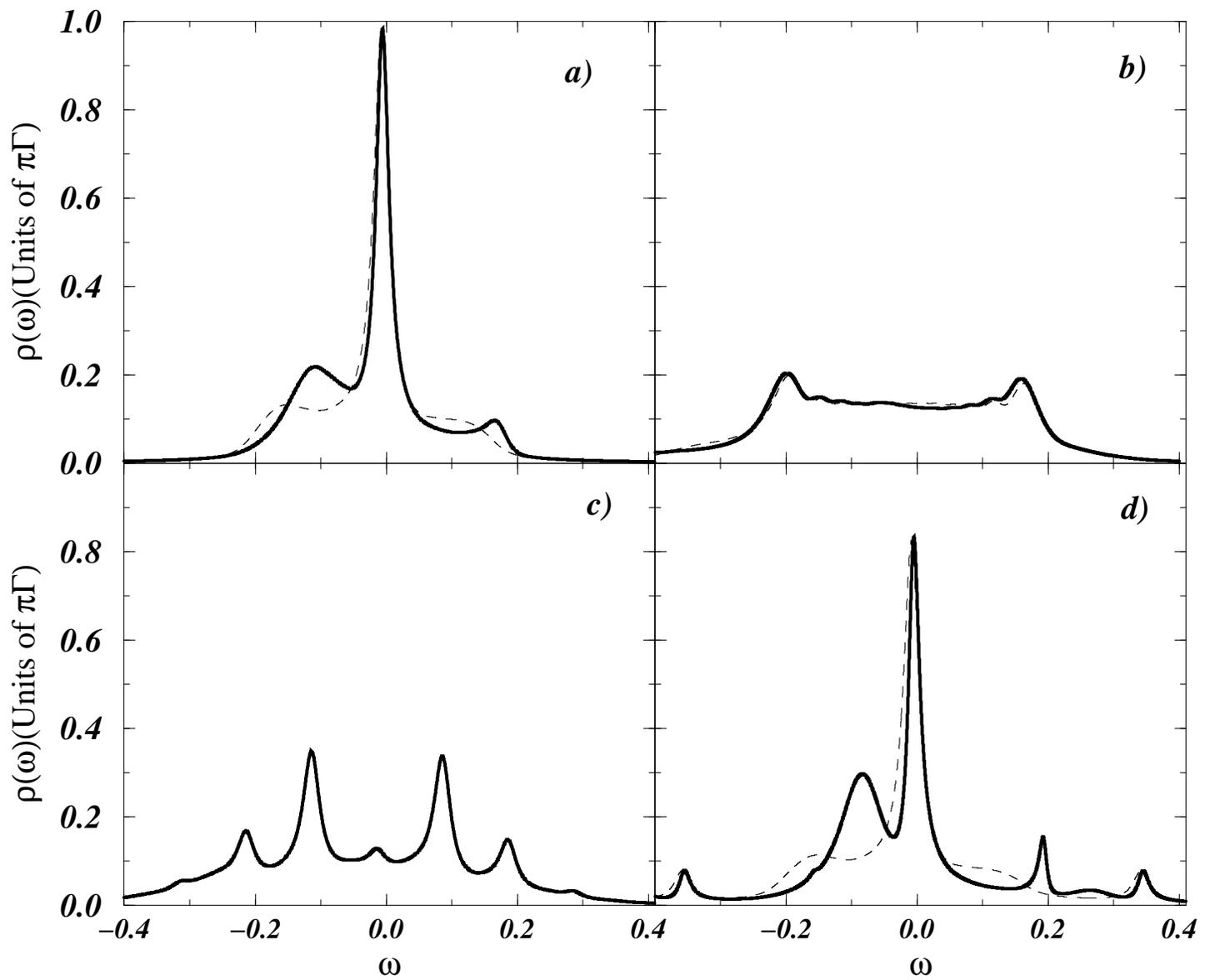
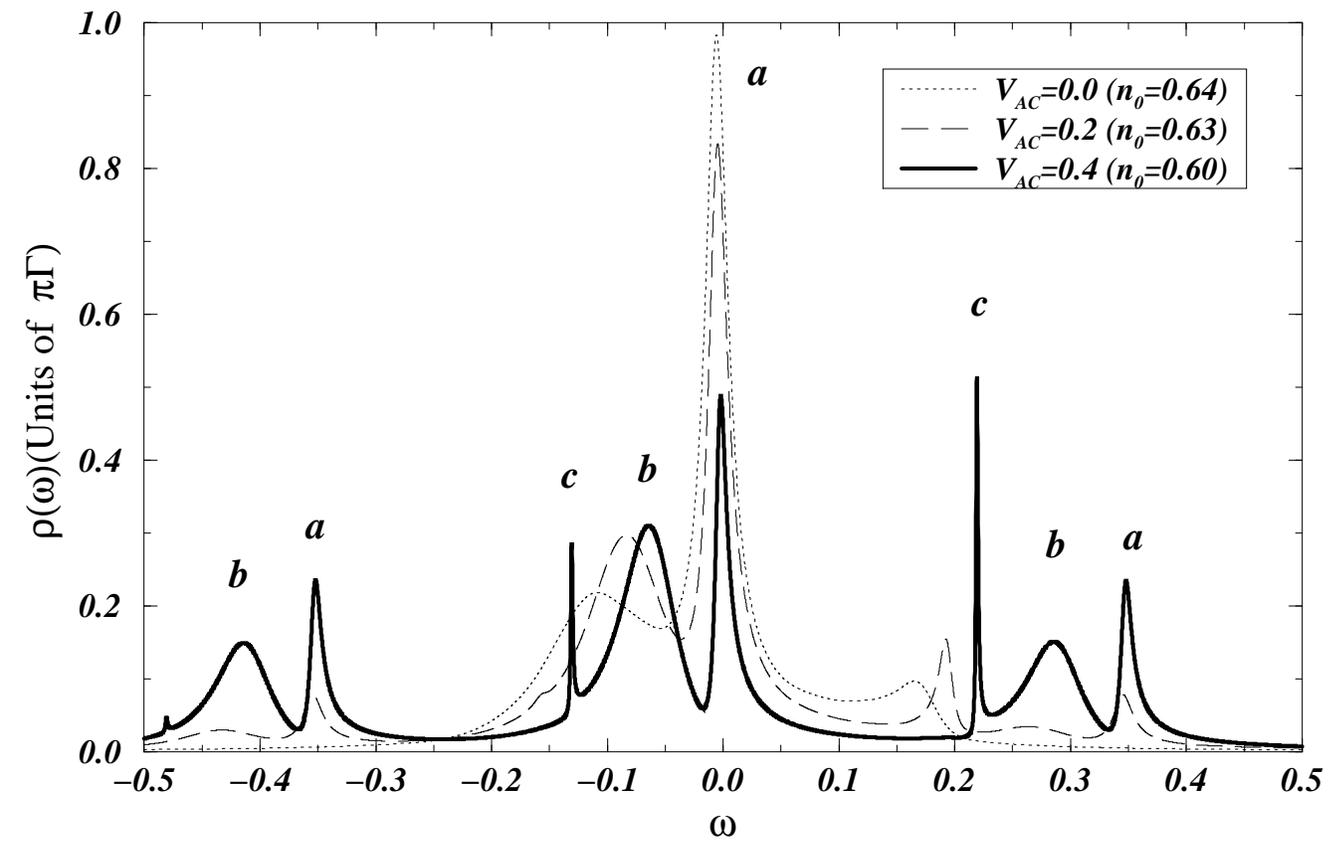
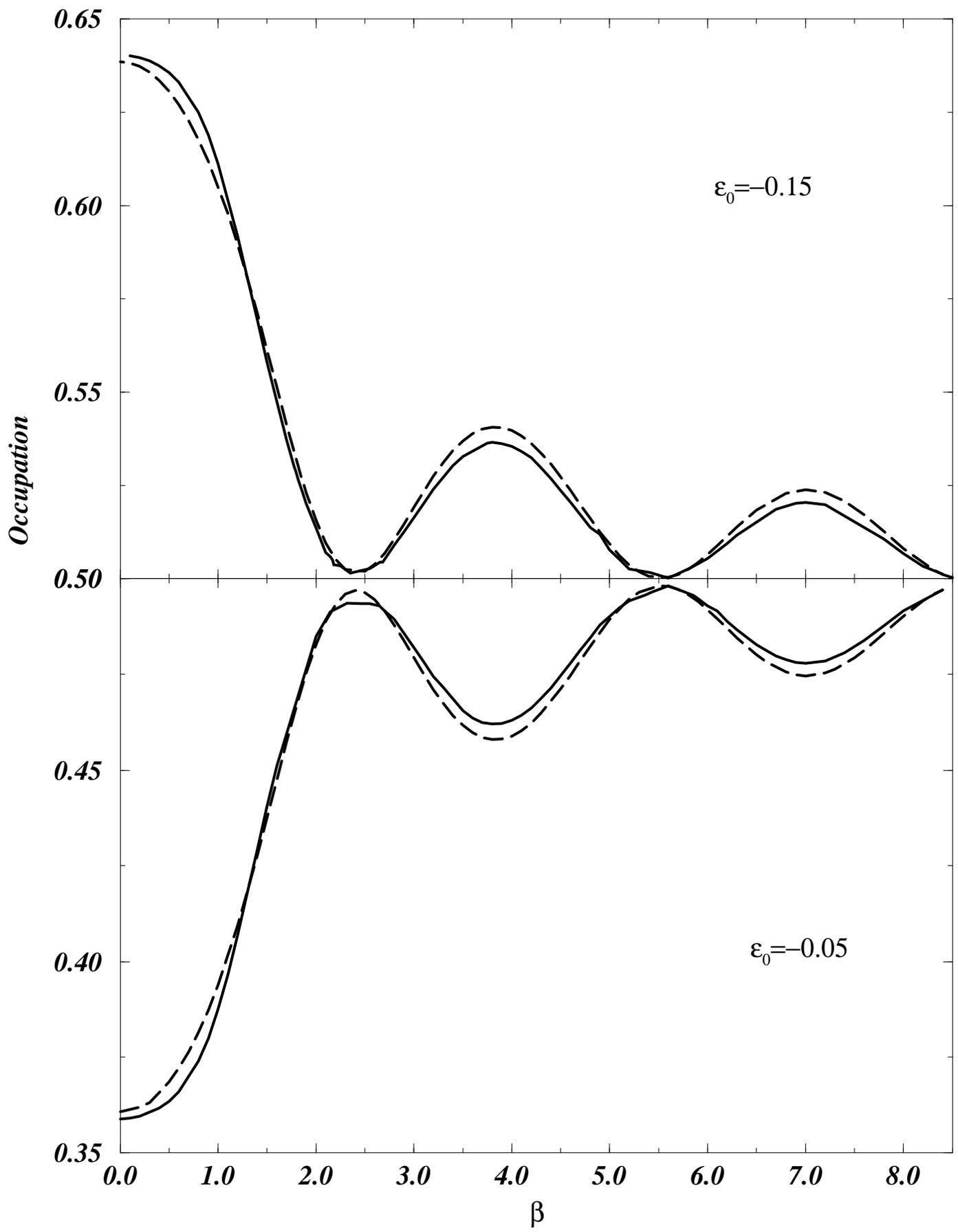
\end{document}